# High pressure effects on the superconductivity of $\beta$-pyrochlore oxides $AOs_2O_6$

T. Muramatsu[1], N. Takeshita[2], C. Terakura[2], H. Takagi[2], Y. Tokura[2], S. Yonezawa[1], Y. Muraoka[1] and Z. Hiroi[1]

[1]*Institute for Solid States Physics, University of Tokyo, Kashiwa, Chiba 277-8581, Japan*

[2]*Correlated Electron Research Center (CERC), National Institute of Advanced Industrial Science and Technology (AIST), Tsukuba, Ibaraki 305-8562, Japan*

abstract

High pressure effects on the superconducting transitions of $\beta$–pyrochlore oxide superconductors $AOs_2O_6$ (A = Cs, Rb, K) are studied by measuring resistivity under high pressures up to 10 GPa. The superconducting transition temperature $T_c$ first increases with increasing pressure in all the compounds and then exhibits a broad maximum at 7.6 K (6 GPa), 8.2 K (2 GPa) and 10 K (0.6 GPa) for A = Cs, Rb and K, respectively. Finally, the superconductivity is suppressed completely at a critical pressure near 7 GPa and 6 GPa for A = Rb and K and probably above 10 GPa for A = Cs. Characteristic changes in the coefficient $A$ of the $T^2$ term in resistivity and residual resistivity are observed, both of which are synchronized with the corresponding change in $T_c$. It is suggested that electron correlations and certain quantum fluctuations play important roles in the occurrence or suppression of superconductivity in the $\beta$-pyrochlore oxides.

A new family of transition metal oxide superconductors $AOs_2O_6$ (A = Cs, Rb, K) called $\beta$-pyrochlore oxides was found recently [1-4]. They exhibit superconductivity at $T_c$ = 3.3 K, 6.3 K and 9.6 K, respectively. Extensive studies are now in progress in order to elucidate the mechanism of the superconductivity. Very recent $\mu$SR [5] and NMR [6] experiments revealed that unconventional superconductivity is realized, particularly in $KOs_2O_6$ with the highest $T_c$. In contrast, there are a few reports insisting on conventional BCS-type superconductivity in $RbOs_2O_6$ [7, 8].

Concerning the normal-state properties, the temperature dependence of resistivity shows $T^2$ behavior at low temperature for every compound, implying that electron correlations are sufficiently strong to stabilize a Fermi liquid state. The coefficient $A$ of the $T^2$ term in resistivity is larger in $KOs_2O_6$ than in $CsOs_2O_6$ and $RbOs_2O_6$. Moreover, it is found that the Sommerfeld coefficient $\gamma$ from specific heat measurements is nearly equal for the Cs and Rb compounds, $\gamma$ = 20 mJ/K$^2$molOs [9], while it is much larger for $KOs_2O_6$ [10]. Hence there must be a large mass enhancement toward $KOs_2O_6$ due to electron correlations.

The superconducting transition temperature of the $\beta$-pyrochlore oxides increases with decreasing the ionic radius of A-site alkali metals from Cs to K. Accordingly, the lattice volume is decreased by the negative chemical pressure. This fact implies that the reduction of the lattice volume by applying physical pressure would further enhance $T_c$. This expectation was confirmed actually in previous two high-pressure (HP) experiments by measuring magnetization up to 1.2 GPa for all the members [11] and up to 0.8 GPa for $RbOs_2O_6$ [7]. Another interesting finding in the former study is that only in the case of $KOs_2O_6$ the $T_c$ exhibits a saturation around 0.6 GPa and then decreases slightly with pressure, suggesting complicated pressure dependences at higher pressures. Thus, it is important to investigate the pressure dependence of $T_c$ and other properties in a wider pressure range. It is also intriguing to search for another ground state which would exist next to the superconducting state under very high pressure, as often found in other unconventional superconductors [12]. In this Letter, we report on resistivity measurements under high pressures up to 10 GPa on all the three members of the $\beta$-pyrochlore oxide superconductors. A characteristic pressure-temperature phase diagram has been obtained, where in general the $T_c$ exhibits a dome-shaped change as a function of pressure.

Polycrystalline samples of $AOs_2O_6$ were prepared as reported previously [1-3]. For $CsOs_2O_6$ and $RbOs_2O_6$, they were nearly single-phase including a small amount of Os metal. For $KOs_2O_6$, 10 mol% of $Cd_2Os_2O_7$ was contained as another phase, which was necessary to obtain a sufficiently hard pellet for resistivity measurements. The coexistence of $Cd_2Os_2O_7$ may not affect the determination of $T_c$, because it becomes insulating below 230 K [13]. Electrical resistivity measurements were performed by the four-probe method in a cubic-anvil press apparatus [14] at high pressures from 2 GPa to 10 GPa and temperatures from 3 K to 300 K. A fluorinert liquid, which is a one-to-one mixture of FC-70 and FC-77 (3M$^{TM}$), was used as a pressure transmitting medium. A polycrystalline pellet of a typical size of 0.8 × 0.3 × 0.4 mm$^3$ was immersed into the fluorinert liquid in a cylindrical teflon capsule of 1.5$^{\phi}$× 1.8 mm$^3$ size. Then, the capsule was put into a cubic pyrophillite block of 6.0 mm edge and compressed almost isostatically by using six anvils made of tungsten carbide. The whole of the sample cell and the press was placed in a liquid-helium dewar to cool down to 3 K. Each run of resistivity measurements was carried out at a constant pressure on cooling and then heating with adjusting the load to the anvils automatically. The load was varied only at room temperature between the measurements. The relation between actual pressures and applying loads had been determined in a separate run by detecting resistivity changes due to the structural phase transitions of Bi, Te and Sn at room temperature.

The temperature dependence of resistivity shows a systematic change under high pressure, as shown in Fig.1. In each compound a drop in resistivity due to superconducting transition is clearly observed at low pressures. The $T_c$ is defined as the midpoint of the transition. The most dramatic change in $T_c$ with pressure is observed in Fig. 1(a) for $CsOs_2O_6$ with the lowest $T_c$ = 3.3 K at AP. The $T_c$ is already raised to 5.0 K at 2 GPa and further increased to 7.6 K at 6 GPa. Then, it turns to decrease at higher pressures and finally reaches to 3.7 K at 10 GPa. It is also to be noted that the resistivity just above $T_c$ varies markedly with pressure: It first decreases and then increases with increasing pressure.

In the case of $RbOs_2O_6$, as shown in Fig.1(b), the $T_c$ at 2 GPa is 8.2 K, raised from 6.3 K at AP. It is reduced gradually with further increasing pressure, and at last no drops in resistivity are observed above 7 GPa down to 4.2 K. Probably, the superconducting transition is suppressed above a critical pressure $P_c$ ~ 7 GPa. In the case of $KOs_2O_6$ with the highest $T_c$ of 9.6 K at AP (Fig. 1(c)), the $T_c$ is reduced to 8.2 K at 2.0 GPa, 5.7 K at 4.0 GPa, and disappears above that. In both cases, the resistivity above $T_c$ is enhanced enormously with increasing pressure.

The pressure dependences of $T_c$ are summarized in Fig.2. Data points below 1.2 GPa were determined by magnetization measurements reported in our previous study [11] and those from 2 to 10 GPa are from the present study. Note that they are connected smoothly to each other. In each compound, $T_c$ initially increases, exhibits a broad maximum, and then decreases with increasing pressure; that is, a dome-shaped variation as a function of pressure. However, the positions of the maximum are quite different among them: ($P^*$/GPa, $T_c^*$/K) = (6.0, 7.6), (2.0, 8.2) and (0.6, 10) for Cs, Rb and K, respectively. Critical pressures $P_c$, where superconductivity is suppressed, are also different; $P_c$ ~ 6 and 7 GPa for K and Rb, and more than 10 GPa for Cs. These differences in $P^*$ and $P_c$ must come from differences in the lattice constant and its compressibility under high pressure. The lattice constant becomes smaller from Cs to K, and the compressibility also becomes smaller in the same direction [15]. Therefore, it is plausible to assume that the $T_c$ of $AOs_2O_6$ exhibits a general dome-shaped dependence on the lattice volume. The enhancement in $T_c$ at AP from Cs to K is ascribed to the smaller lattice volume. On the other hand, the observed difference in $T_c^*$ among the three compounds must come from another degree of freedom. Possibly, it is related to the only one crystallographic parameter of the $\beta$-pyrochlore structure, which is the $x$ parameter of the 48$f$ oxygen. This parameter determines the magnitude of a trigonal distortion for $OsO_6$ octahedra and thus may influence the band structures of the $\beta$-pyrochlroe oxides.

Among the three compounds, $RbOs_2O_6$ would provide us with a suitable playground to study the pressure dependence of resistivity in detail, because the entire range of the $T_c$ dome, as well as a non-superconducting regime, can be examined in one system. The temperature dependences of resistivity for $RbOs_2O_6$ under various pressures are shown in Fig. 3. In the pressure range below 4 GPa (Fig. 3a), where $T_c$ exhibits a broad maximum, the overall temperature dependence of resistivity is rather similar to each other. Resistivity near room temperature decreases with pressure, while that at low temperature above $T_c$ increases slightly. This reduction of resistivity near room temperature probably is not intrinsic but may come from the compression of grains in the polycrystalline sample used in this work. In contrast, the enhancement of the low temperature resistivity must be intrinsic and implies that the residual resistivity is increased with pressure.

On the other hand, in the pressure range between 4 and 7 GPa which corresponds to the right half of the $T_c$ dome, a remarkable change in resistivity is observed as shown in Fig. 3(b). The room temperature resistivity becomes now almost independent of pressure, while the residual resistivity is further increased enormously. As a result, the resistivity at 7 GPa becomes almost temperature independent. Applying further pressure, the overall resistivity shifts upward uniformly at 8 GPa, and surprisingly at 10 GPa the resistivity starts to decrease steeply below 100 K, implying that metallic behavior is recovered. However, it is to be noted that this temperature dependence of resistivity is apparently different from that at low pressure in the superconducting regime. Very recently, we carried out another HP experiments on a different sample of $RbOs_2O_6$ with better purity and larger residual resistivity ratio and obtained essentially the same results, indicating that the above changes are intrinsic and reproducible. We found substantially the same behavior in resistivity under HP for $CsOs_2O_6$ and $KOs_2O_6$, except for differences in the pressure range.

It is known that the low-temperature resistivity of the $\beta$-pyrochlore oxides generally exhibits $T^2$ behavior, indicative of large electron correlations. The coefficient $A$ of the $T^2$ term is much larger in $KOs_2O_6$ than in $CsOs_2O_6$ and $RbOs_2O_6$, suggesting that electron correlations play an important role in the enhancement of $T_c$. Here, we demonstrate how pressure affects $A$ as well as residual resistivity $\rho_0$. Resistivity at each pressure is plotted against $T^2$, as shown in the inset to Fig.3(a) for $RbOs_2O_6$, to determine $A$ and $\rho_0$. The slope of the $T^2$ term in $RbOs_2O_6$ is large at 2 GPa and does not change much up to 3 GPa near the $T_c$ maximum. Then it decreases very much from 4 GPa to 7 GPa, where $T_c$ decreases and finally disappears, and then increases again at 8 GPa and 10 GPa. The origin of an upturn observed at low temperatures above 7 GPa is not known and it is ignored in evaluating $A$ and $\rho_0$. Such an upturn was not detected in our recent experiment using a better quality of sample. The same analysis has been carried out for $CsOs_2O_6$. The resulting pressure dependences of $A$ and $\rho_0$ are summarized in Fig. 4.

In order to deduce a general trend, each quantity has been normalized by the value at the maximum of the $T_c$ dome. Pressure is also normalized as $P_n = (P - P^*)/(P_c - P^*)$, where $P^*$ is a pressure at the $T_c$ maximum and $P_c$ is a critical pressure where superconductivity disappears; that is, $P_n = 0$ for $P = P^*$ and $P_n = 1$ for $P = P_c$. For $KOs_2O_6$, only $T_c$ is plotted, because we could not obtain reliable data due to the presence of impurities. It becomes apparent now that normalized $T_c$ falls on a universal curve. Either normalized $A$ and $\rho_0$ is connected smoothly between $RbOs_2O_6$ and $CsOs_2O_6$ to show a general curve: Normalized $A$ exhibits a broad maximum near the $T_c$ maximum and a minimum near $P_c$, while normalized $\rho_0$ shows a minimum near $P^*$ and a maximum near $P_c$. In the framework of Fermi liquid theory, $A$ is proportional to square of effective mass of carriers or density of states at the Fermi level. Thus, the above results on $A$ imply that the carrier mass is enhanced significantly at the $T_c$ maximum owing to correlation effects. This strongly suggests that the mechanism of superconductivity in the $\beta$-pyrochlore oxides is relevant to electron correlations.

On the other hand, residual resistivity is generally expressed by the equation $\rho_0 = (\hbar / e^2 l)(3\pi^2)^{1/3} n^{-2/3}$, where $l$ is the mean free path of carriers and $n$ is carrier density. The mean free path at low temperature may be determined by the concentration of impurities or defects and thus should not be affected by pressure. It is expected, however, that the mean free path would be reduced largely by quantum fluctuations developing toward absolute zero, which often exist near a quantum critical point associated with a certain long-range order such as a magnetic transition or valence transition [16,17]. The observed enhancement in $\rho_0$ near $P_c$ may indicate that there is an unknown long-range order at higher pressure. Possibly, the sudden recovery of metallic behavior observed below 100 K at 10 GPa in $RbOs_2O_6$ (Fig.3(b)) is related to this long range order. It is plausible that the associated quantum fluctuations would suppress the superconductivity, giving rise to a dome-shaped $T_c$ variation.

The observed dome-shaped change in $T_c$ and the corresponding unusual variations in $A$ and $\rho_0$ must reflect unconventional features of superconductivity in the $\beta$-pyrochlore oxides. However, it seems that they are quite different from those observed in well-known unconventional superconductors, such as $f$-electron intermetallic compounds or low-dimensional organic conductors [12,18]. A typical $f$-electron compound $CeCu_2Ge_2$ exhibits a large reduction in $A$ and a sharp peak in $\rho_0$ at the $T_c$ maximum, which is believed to be induced by valence fluctuations [12,16,17]. In the present $\beta$-pyrochlores osmium ions are formally in a 5.5 valent state with two and a half 5$d$ electrons. It is interesting to note that this number of 5$d$ electrons is just between two integer values for $Re^{5+}$ ions (5$d^2$) in 1 K superconductor $Cd_2Re_2O_7$ and $Os^{5+}$ ions (5$d^3$) in $Cd_2Os_2O_7$ exhibiting a metal-to-insulator transition [13, 19]. However, we think that valence fluctuations may not play a primary role in the $\beta$-pyrochlores, because the band width of Os 5$d$ – O 2$p$ hybridized bands is large, ~3 eV, from the band structure calculations [20-22]. The most plausible scenario for quantum fluctuations is to assume electronic instability associated with Fermi surface nesting. Harima has pointed out that an important general feature on the band structure of the $\beta$-pyrochlores is the existence of a pair of Fermi surfaces of nearly octahedral shape which forms a thin shell centered at the $\Gamma$ point. Thus, strong Fermi surface nesting and resulting spin-density-wave instability are expected. Moreover, a 'dimple' grows on the octahedral Fermi surface from K to Cs, suggesting that the nesting becomes weaker along this line. It would be interesting to test how this feature changes under high pressure. In order to clarify the existence of possible quantum fluctuations in the $\beta$-pyrochlore oxides and their relevance to the superconductivity, further experiments are required under high pressures.

In summary, we have obtained interesting and characteristic behavior in the pressure dependence of $T_c$ and other parameters in the $\beta$-pyrochlore oxides $AOs_2O_6$ by measuring resistivity under high pressure up to 10 GPa. Generally, the $T_c$ shows a dome-like shape as a function of pressure. The variations of the coefficient $A$ of the $T^2$ term and the residual resistivity $\rho_0$ are also anomalous and synchronized with the change in $T_c$. These results suggest that electron correlations enhance the superconductivity, while certain quantum fluctuations give rise to the degradation of $T_c$ at higher pressure.

This research was supported by a Grant-in-Aid for Scientific Research B (16340101) provided by the Ministry of Education, Culture, Sports, Science and Technology, Japan. T. M. thanks for financial support by Grant-in-Aid for Creative Scientific Research (13NP0201).

Reference

[1] S. Yonezawa, Y. Muraoka, Y. Matsushita, and Z. Hiroi, J. Phys.:Condens. Matter **16**, L9 (2004).

[2] S. Yonezawa, Y. Muraoka, Y. Matsushita, and Z. Hiroi, J. Phys. Soc. Jpn. **73,** 819 (2004).

[3] S. Yonezawa, Y. Muraoka, and Z. Hiroi, J. Phys. Soc. Jpn. **73,** 1655 (2004).

[4] Z. Hiroi, S. Yonezawa, and Y. Muraoka, J. Phys. Soc. Jpn. **73,** 1651 (2004).

[5] A. Koda, W. Higemoto, K. Ohishi, S. R. Saha, R. Kadono, S. Yonezawa, Y. Muraoka, and Z. Hiroi, cond-mat/0402400.

[6] K. Arai, J. Kikuchi, K. Kodama, M. Takigawa, S. Yonezawa, Y. Muraoka, and Z. Hiroi, Proceedings of SCES'04 (2004).

[7] R. Khasanov, D. G. Eshchenko, J. Karpinski, S. M. Kazakov, N. D. Zhigadlo, R. Brütsch, D. Gavillet, D. Di Castro, A. Shengelaya, F. La Mattina, A. Maisuradze, C. Baines, and H. Keller, Phys. Rev. Lett. **93,** 157004 (2004).

[8] M. Brühwiler, S. M. Kazakov, N. D. Zhigadlo, J. Karpinski, and B. Batlogg, Phys. Rev. B **70**, 020503(R) (2004).

[9] Z. Hiroi, S. Yonezawa, T. Muramatsu, J. Yamaura, and Y. Muraoka, J. Phys. Soc. Jpn, in press.

[10] Z. Hiroi, S. Yonezawa, J. Yamaura, T. Muramatsu, and Y. Muraoka, submitted to J. Phys. Soc. Jpn.

[11] T. Muramatsu, S. Yonezawa, Y. Muraoka, and Z. Hiroi, J. Phys. Soc. Jpn. **73** 2912 (2004).

[12] D. Jaccard, H. Wilhelm, K. Alami-Yadri, and E. Vargoz, Physca B **259-161** 1 (1999).

[13] A. W. Sleight, J. L. Gillson, J. F. Weiher, and W. Bindloss, Solid State Commun. **14** 357 (1974).

[14] N. Mori, *et. al*. Jpn. J. Appl. Phys. Ser. **8**, 128 (1993).

[15] J. Yamaura, private communication.

[16] K. Miyake, O. Narikiyo, and Y. Onishi, Physica B **259-261** 676 (1999).

[17] Y. Onishi, and K. Miyake, J. Phys. Soc. Jpn. **69** 3955 (2000).

[18] D. Jerome, Science **252** 1509 (1991).

[19] D. Mandrus, J. R. Thompson, R. Gaal, L. Forro, J. C. Bryan, B. C. Chakoumakos, L. M. Woods, B. C. Sales, R. S. Fishman and V. Keppens, Phys. Rev. B **63** 195104 (2001).

[20] R. Saniz, J. E. Medvedeva, Lin-Hui Ye, T. Shishidou, and A. J. Freeman, Phys. Rev. B **70** 100505(R) (2004)

[21] J. Kunes, T. Jeong and W. E. Pickett, Phys. Rev. B **70** 174510 (2004)

[22] H. Harima, in preparation.

Figure captions

Fig. 1

Superconducting transitions of $\beta$-pyrochlore oxides $AOs_2O_6$ (A = Cs, Rb, K) detected by resistivity measurements under high pressures from 2 GPa to 10 GPa. The $T_c$ at ambient pressure is 3.3 K for Cs, 6.3 K for Rb and 9.6 K for K.

Fig. 2

Pressure dependence of superconducting transition temperature $T_c$ for the three $\beta$-pyrochlore oxides. Open marks below 1.2 GPa are the data from previous magnetization measurements [11] and solid marks represent $T_c$ determined in the present resistivity measurements. Squares, triangles and circles correspond to Cs, Rb and K, respectively. Solid and broken lines are guides to the eye.

Fig. 3

The evolution of temperature dependence of resistivity $\rho$ for $RbOs_2O_6$ in the pressure ranges of 2 – 4 GPa (a) and 4 – 10 GPa (b). $\rho$ - $\rho_0$ is plotted against $T^2$ in the inset of (b), where $\rho_0$ is residual resistivity.

Fig. 4

Pressure dependence of $T_c$ (a), coefficient $A$ (b), and residual resistivity $\rho_0$ (c). In order to deduce general features, normalized pressure $P_n$ is used, which is given by the equation $P_n = ( P - P^* ) / ( P_c - P^* )$, where $P^*$ and $P_c$ is pressures with the maximum $T_c$ and $T_c \sim 0$, respectively. $A$ and $\rho_0$ are also normalized by the corresponding values at the maximum $T_c$. Squares, triangles and circles represent data for A = Cs, Rb and K, respectively.

Fig. 1

(a)

(b)

(c)

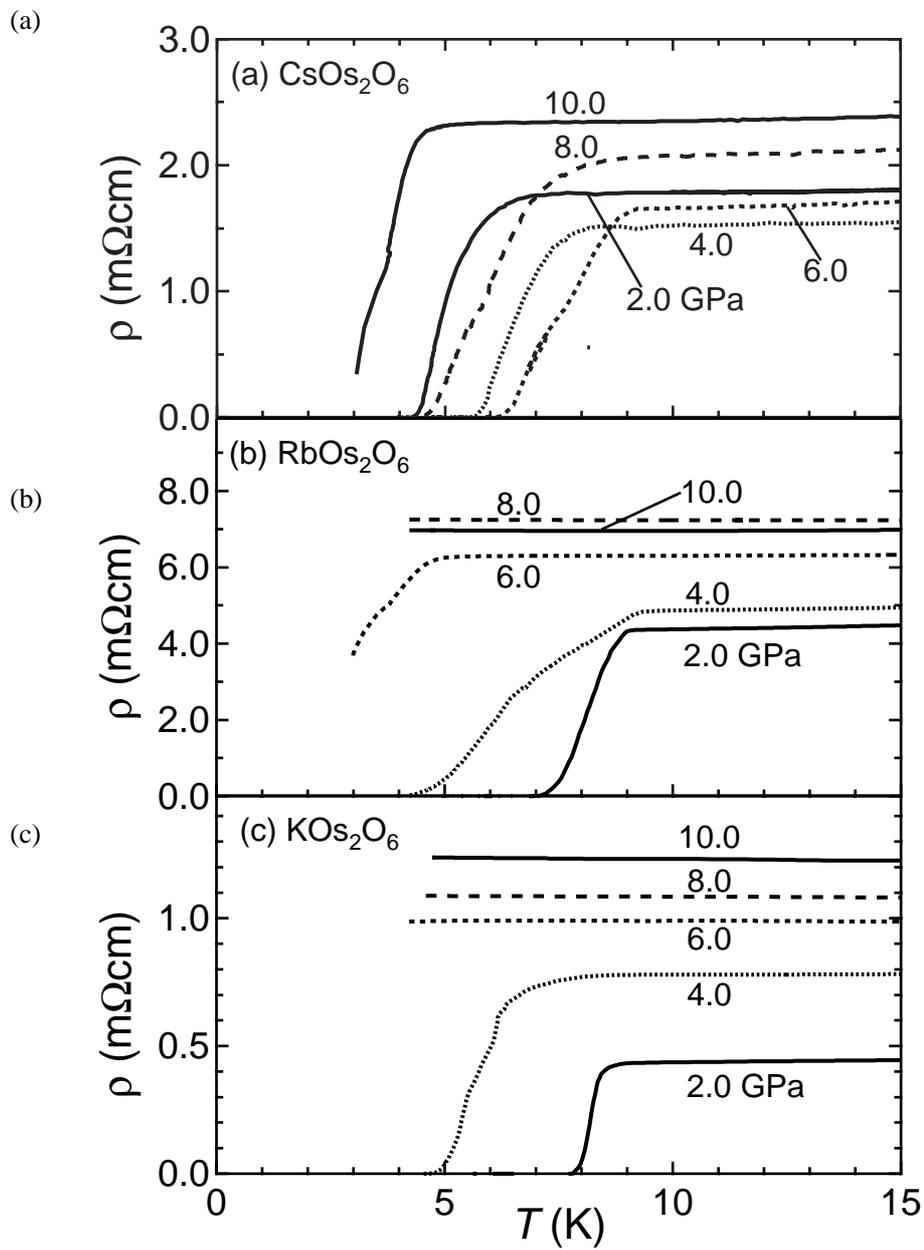

Fig. 2

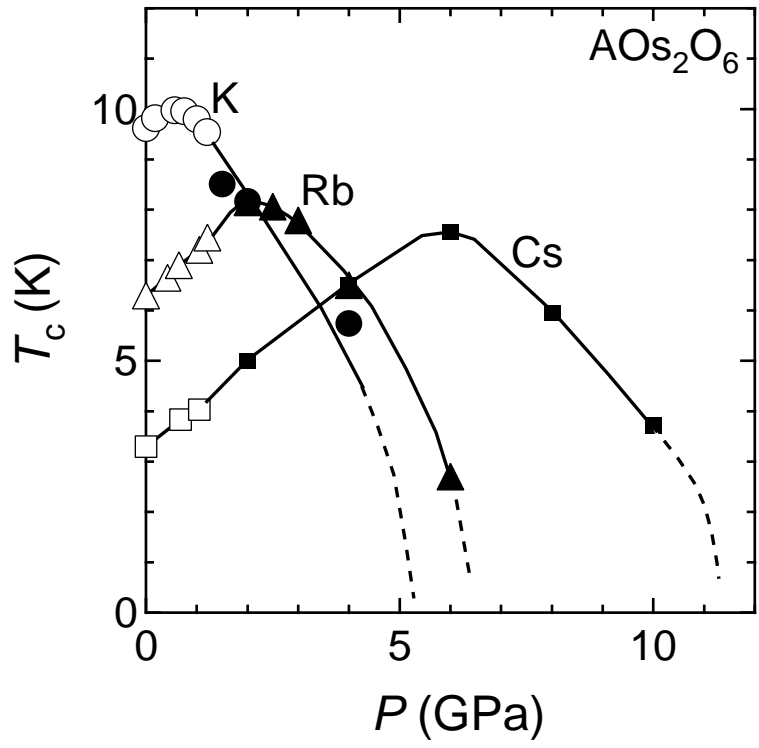

Fig. 3

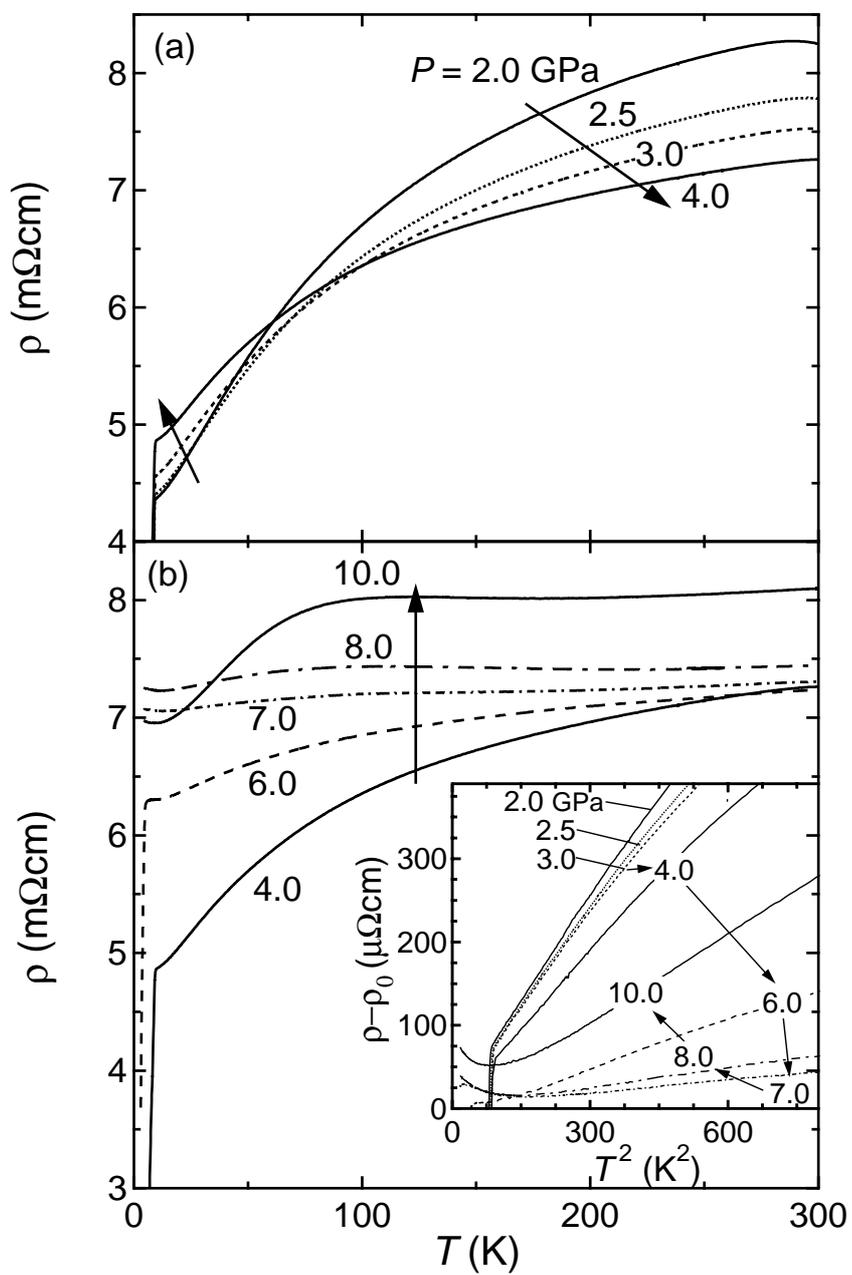

Fig. 4

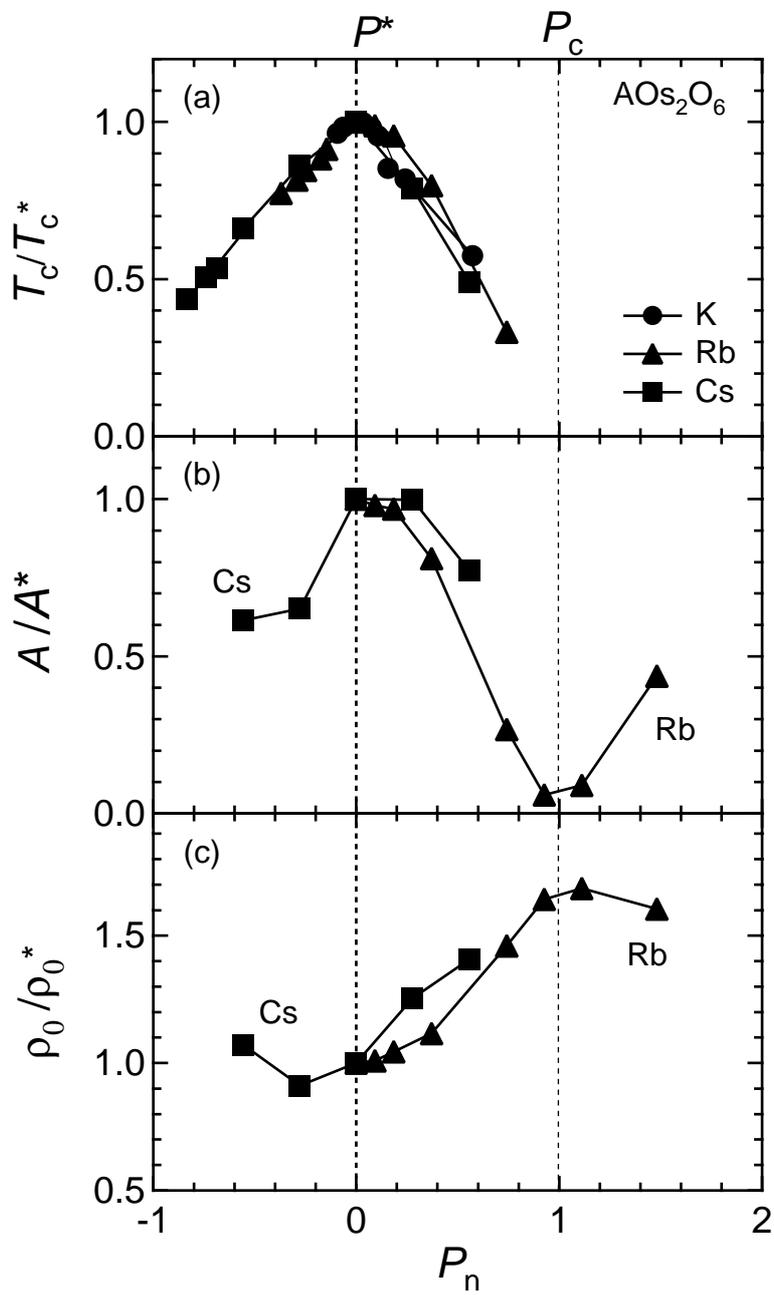